\documentclass[12pt,preprint]{aastex}

\newcommand{\degree}{$^o$}
\def\msun{\hbox{\rm M$_{\scriptsize \odot}$}}

\shorttitle{Spitzer's view on an outer Galaxy dark cloud}
\shortauthors{Frieswijk et al.}

\begin{document}

\title{Spitzer's mid-infrared view on an outer Galaxy Infrared Dark 
Cloud candidate toward NGC 7538}
\author{W. F. Frieswijk\altaffilmark{1,2}, M. Spaans\altaffilmark{1},
R. F. Shipman\altaffilmark{2},\\D. Teyssier\altaffilmark{3}, 
S. J. Carey\altaffilmark{4} and A. G. G. M. Tielens\altaffilmark{5}}
\altaffiltext{1}{Kapteyn Astronomical Institute, P.O. Box 800, 
9700 AV Groningen, The Netherlands}
\altaffiltext{2}{National Institute for Space Research, P.O. Box 800,
9700 AV Groningen, The Netherlands}
\altaffiltext{3}{ESAC, Urb. 
Villafranca del Castillo, P.O. Box 50727, Madrid 28080, Spain}
\altaffiltext{4}{SSC, California Institute of
Technology, MC314-6, 1200 East California Boulevard, Pasadena, CA 91125}
\altaffiltext{5}{NASA Ames Research Center, MS 245-3, Moffett 
Field, CA 9435-1000, USA}
\altaffiltext{}{Electronic adress: frieswyk@astro.rug.nl, spaans@astro.rug.nl,
R.F.Shipman@sron.nl,\\ David.Teyssier@sciops.esa.int, carey@ipac.caltech.edu,
Alexander.G.Tielens@nasa.gov}
\begin{abstract}
	Infrared Dark Clouds (IRDCs) represent the earliest observed stages of 
	clustered star formation, characterized by large column densities of cold 
	and dense molecular material observed in silhouette against a bright 
	background of mid-IR emission. Up to now, IRDCs were predominantly known 
	toward the inner Galaxy where background infrared emission levels are high.
	We present {\it Spitzer} observations 
	with the Infrared Camera Array toward object G111.80+0.58 (G111) in the outer 
	Galactic Plane, located at a distance of $\sim$3\,kpc from us and $\sim$10\,kpc 
	from the Galactic center. Earlier results show that G111 is a massive, 
	cold molecular clump very similar to IRDCs.
	The mid-IR {\it Spitzer} observations unambiguously detect object G111 
	in absorption. We have identified for the first time an IRDC 
	in the outer Galaxy, which confirms the suggestion that cluster-forming
	clumps 
	are present throughout the Galactic Plane. However, against a low mid-IR background 
	such as the outer Galaxy it takes some effort to find them.
\end{abstract}
\section{Introduction}
Massive stars are believed to form almost exclusively in stellar clusters 
\citep{Blaauw1964,deWit2005}, where supposedly the majority of all stars in 
the Galaxy form. The precursors to these stellar clusters are cold, dense and, 
most importantly, massive molecular clumps. 
A major, unexpected and exciting result from the {\it Midcourse Space Experiment} 
({\it MSX}) and {\it Infrared Space Observatory} ({\it ISO}) is the discovery 
of dark clouds seen in silhouette against the bright mid-IR background 
toward the inner Galactic Plane \citep{MSXIRDC1998,ISOCAM1996}. In contrast 
to the well-known dark clouds seen in visual extinction against the stellar 
distribution, these so-called Infrared Dark Clouds (IRDCs) have much higher 
column densities ($\gtrsim$\,10$^{23}$\,cm$^{-2}$), are 
at large 
distances ($\gtrsim$\,1\,kpc) and hence are typically much more massive.
Follow-up studies \citep[e.g.][]{Carey1998,Teyssier2002,Simon2006co} 
show that they are cold ($<$\,20\,K), dense ($\sim$\,10$^5$\,cm$^{-3}$) and, 
indeed, among the most massive molecular clumps yet found in our Galaxy
($M$$\sim$100--10$^5$\,M$_{\odot}$).
Many IRDCs contain compact (sub-) millimeter cores 
\citep[$M$$\sim$10--10$^3$\,\msun, $n$$\sim$10$^3$--10$^7$\,cm$^{-3}$, 
$T$$\sim$15--30\,K, e.g.,][]{Carey2000,Teyssier2002,Garay2004,Rathborne2006}. 
While these cores were first considered to represent the pre-stellar 
core phase of clustered star formation, recent studies show that at least
some cores in IRDCs contain 
proto-stellar objects 
\citep[e.g.,][]{Redman2003,Ormel2005,Rathborne2005}. 
Nevertheless, IRDCs
are likely to represent the earliest stages of clustered star formation 
\citep[e.g.,][]{Menten2005,Rathborne2006} and they are generally referred to as
cluster-forming clumps. Their ambient physical conditions 
can illuminate the role of IRDCs in the star forming process and can provide 
insight into the differences between low- and high-mass star formation, the 
nature of the initial mass function and the impact of environment on star 
formation. Moreover, their 
core forming properties may 
reveal the important mechanisms involved in forming high-mass stars, e.g., 
competitive accretion, merging of lower-mass stars or massive disk accretion 
\citep[e.g.,][and references therein]{Larson2007}.

The identification of IRDCs is by necessity biased toward cold molecular 
clouds having sufficient column density against a bright mid-IR background,
in order to be seen as absorption features. Hence, identifying IRDCs 
is considerably easier toward the mid-IR bright spiral arms and molecular 
ring in the inner Galactic Plane \citep{Simon2006msx}, i.e., within 
90\degree\ from the Galactic Center.
An unbiased, more complete understanding of clustered star formation and 
its dependence on environment and other external properties, e.g., 
interstellar radiation field, metallicity, external pressure and dynamics,
requires a study of IRDC-like objects in more quiescent regions, such as 
the outer Galaxy, i.e., between $l$=90\degree and $l$=270\degree.

Frieswijk et al. (2008) 
produce a catalog of extended red objects in the outer Galactic Plane.
The objects are identified in the {\it Two Micron All Sky Survey Point 
Source Catalog} ({\it 2MASS,} \citealt{Skrutskie2006}) as clusters of stars 
that show a statistically redder color distribution compared to their local
surroundings.
A sample, consisting of 414 objects covered by the Canadian Galactic Plane 
Survey \citep[CGPS,][]{Heyer1998}, was investigated in detail using the
CO data of the CGPS. Over 90\% of the objects correlate morphologically with CO
emission with a typical FWHM of the order of 5\,km\,s$^{-1}$.
This suggests that the observed reddening is due to the presence of foreground 
molecular clouds. The kinematic distance of the clouds, derived from the 
CO radial velocity, indicates that $\sim$15\% are located beyond 
$\sim$\,3\,kpc from us. Some of these distant clouds ($\sim$10) 
do not show significant mid- and far-IR emission in, e.g., MSX and IRAS,
and are potentially cold, cluster-forming clumps in the outer Galaxy, 
or `IRDC-like' objects.
\citet[][F07]{Frieswijk2007AA} studied the physical 
characteristics of a candidate IRDC-like object (G111) by using molecular 
line observations. At a radial velocity of -52\,km\,s$^{-1}$, G111 is 
located at a kinematic distance of $\sim$\,5\,kpc, assuming IAU standard constants. 
However, within 15\arcmin\ on the sky and at similar radial velocity are
the (massive) star forming regions NGC 7538 and S 159. These are part of the Cas 
OB2 complex in the Perseus spiral arm, located at $\sim$3\,kpc from us 
and $\sim$10\,kpc from the Galactic center. Hence, a distance of 3\,kpc is 
assumed for G111.
A detailed investigation of the molecular lines conducted by F07 revealed multiple 
cold and massive cores (P1--P10, see Fig.\ref{fig1} and Sec.\ref{conclusions})
along a filamentary cloud structure. The main conclusion is that G111 has 
global properties in agreement with values found for IRDCs, 
e.g., $M$\,$\gtrsim$\,3000\,\msun, 
$N$\,$\gtrsim$\,10$^{23}$\,cm$^{-2}$, $T$\,$\lesssim$\,20\,K.
The mass and density of individual cores are low when compared to sub-millimeter cores 
in IRDCs.
However, with peak column densities in excess of 10$^{23}$\,cm$^{-2}$, 
the most massive cores in G111 should, provided that sufficient mid-IR background 
is present, be observable as absorption features in a similar way to inner 
Galaxy IRDC cores.

We subsequently requested {\it Spitzer} observations toward G111 
to test the idea that the cores can be observed in absorption. {\it MSX} 
observations were not able to resolve this, presumably because the low 
sensitivity did not allow a detection of contrast between the molecular cloud 
and the background. With the sensitivity of the Infrared Camera Array (IRAC) on 
{\it Spitzer} we argued that it should be possible to observe this contrast. 
In this paper, we present the 
first outcomes of the data. Combined with the results in F07, we conclude that 
we have, beyond doubt, a detection of an IRDC in the outer Galaxy.
The paper is organized as follows: Section \ref{observations} gives 
an
overview of the observations. Section \ref{results} presents the results. In 
Section \ref{conclusions} we discuss the observations and give our interpretation 
of the data 
compared to the results published in F07. We conclude the section with some 
future prospects.
\section{Observations}
\label{observations}
The data presented in this paper are part of a larger project that focusses 
on the star formation characteristics of cluster-forming clumps in the outer Galaxy. 
The 3.5--8\,$\mu$m observations were carried out on October 23, 2007 using the IRAC 
instrument on the {\it Spitzer Space Telescope}. Additional IRAC data are expected,
as well as 24\,$\mu$m data from the Multiband Imaging Photometer (MIPS) on {\it Spitzer}. 
These will be used to analyze the proto-stellar 
content of the region. However, the emphasis of this letter is to present the 
first identification of an IRDC in the outer Galaxy. For that purpose, the present 
data are adequate.

The IRAC observations consist of a cycling 7-point dither pattern with medium scale 
step-size in high dynamic range mode. 
The data were processed and transformed into mosaic images using the data processing 
pipeline version S16.1.0 provided by the {\it Spitzer} Science Center. 
An integration time of 10.4\,s per pointing resulted in an rms-noise at 8\,$\mu$m of 
$\sim$0.05\,MJy\,sr$^{-1}$ and a point source sensitivity of $\sim$10\,$\mu$Jy. Additional 
calibration of the data is certainly required for the analysis of band-merged 
images. However, the standard pipeline produces data of sufficient quality to examine 
the contrast at 8\,$\mu$m presented in this paper.
\section{Results}
\label{results}
Figure \ref{fig1} shows the 8\,$\mu$m emission toward G111. 
The image reveals a dark, filamentary 
structure against a bright and diffuse mid-IR background. The diffuse 
emission resembles the feathery cirrus emission first observed with the 
{\it Infrared Astronomical Satellite} \citep[{\it IRAS}, e.g.,][]{Low1984}. 
The cirrus observed with {\it IRAS} is assumed to be predominantly local, 
but here at least part of the emission appears to be located behind G111
($\gtrsim$\,3\,kpc).

The contours show the integrated C$^{18}$O 2-1 emission observed by F07. 
The dark structures match the contours of the C$^{18}$O strikingly well. 
This implies that we observe a cold, dense molecular cloud, unlike the 
dark `hole' appearing at the right side of the image, where we do not see a 
match with C$^{18}$O. The locations in the figure indicated by the numbers P1 
to P10 represent target positions observed in detail by F07.
The positions were selected on basis of their high column density. The physical
properties derived by F07 suggest that they represent cold and dense molecular cores 
where stars might form. We briefly summarize their properties and relate these with 
the current observations in Section \ref{conclusions}.

The 8\,$\mu$m image reveals several clusters of point-like objects along the 
cloud filaments, most noticeably near P5 and P8. The upper two panels on the 
right 
of Figure \ref{fig1} display false-color 
close-ups of these 
positions in the four IRAC bands, i.e., 3.6\,$\mu$m (blue), 4.5\,$\mu$m (green), 
5.8\,$\mu$m (yellow) and 8\,$\mu$m (red). 
The lower panel displays 
a 
close-up of the bright source IRAS23136+6111. The intensities there 
are scaled 
so that the dark lane, corresponding to the 
C$^{18}$O filament,
stands out clearly against the 
emission of the IRAS source.

We quantify the correlation between the dark features observed in the 
{\it Spitzer} data and the integrated C$^{18}$O intensity
by determining the mean 
8\,$\mu$m diffuse emission in increasing C$^{18}$O emission bins. 
In order to remove the contribution of bright point sources, we first smooth 
the 8\,$\mu$m image with a median filter of 11 pixels wide (1 pixel = 1.2\arcsec). 
We choose a C$^{18}$O bin-size of 1\,K\,km\,s$^{-1}$ and use only the region that 
is actually covered by the C$^{18}$O observations. The results are given 
schematically in Figure \ref{fig2}. The first bin in the diagram represents an 
estimate of the background, determined from several positions in the image away
from the dark cloud, after filtering as described above. An alternative background 
estimate is provided by the second bin in the diagram, which represents the 
8\,$\mu$m diffuse emission at any position where the C$^{18}$O emission is less
than 1\,K\,km\,s$^{-1}$. The last bin of the diagram represents the lowest 
8\,$\mu$m emission toward locations where there is C$^{18}$O emission in excess of 
1\,K\,km\,s$^{-1}$. The dashed boxes for each bin represent the scatter of the 
8\,$\mu$m emission. Note that the scatter is significant in the lower 
C$^{18}$O bins (1--4\,K\,km\,s$^{-1}$) because the extended, bright 
8\,$\mu$m emission falls within these contours (see Figure \ref{fig1}).

If we adopt a background emission of 21\,MJy\,sr$^{-1}$, and assume that the 
low emission, i.e., 13\,MJy\,sr$^{-1}$, is purely due to absorption by the 
molecular cloud, then we can estimate the extinction. Using standard conversions 
given in the IRAC data handbook, the decrease in 8\,$\mu$m emission of 
8\,MJy\,sr$^{-1}$ corresponds to 0.5\,mag of extinction.
This value represents only a lower limit. Both the cloud and the background are 
contaminated by emission from foreground material and zodiacal light. The 
contribution of the zodiacal light is 4.2\,MJy\,sr$^{-1}$, adopted from the 
{\it Spitzer} observation planning tool (SPOT). The foreground is difficult to 
determine. If we conservatively (remember the cloud is at 3\,kpc) assume a 
foreground emission contribution of 4\,MJy\,sr$^{-1}$, then the magnitude of the 
extinction at 8\,$\mu$m is unity. For convenience, the mid-IR extinction can be 
expressed in terms of visual extinction using 
$A_{\rm{V}} \sim 25\times A_{8\mu\rm{m}}$, where a standard conversion is adopted 
from \citet{Rieke1985}. The arrow in Figure \ref{fig2} corresponds to an 
$A_{\rm{V}}$ of 10\,mag assuming a background of 21\,MJy\,sr$^{-1}$.
\section{Conclusions}
\label{conclusions}
The mid-IR properties of G111 compare well to other IRDCs.
The {\it MSX} survey data have been used by \citet{Simon2006msx} to identify a large number 
of IRDCs in the inner Galaxy by their mid-IR contrast relative to the background.
A typical 8\,$\mu$m extinction observed toward IRDCs, determined from the ratio 
between the on- and off-source emission after foreground and zodiacal light correction, 
is 1--2\,mag \citep[e.g.,][]{MSXIRDC1998,Carey2000}. We find extinction values
toward G111 to be in agreement with this, at least toward the darkest cores.
More so, the extinction measurement gives only a strict lower limit to the column of 
material. The cloud does not only absorb background emission, but re-radiates 
low-level emission as well. For a cold dark cloud ($<$25\,K) this will be negligible. 
However, once embedded objects start heating parts of the cloud it may become important. 
The contrast between the cloud and the background, and thus the measured extinction, 
decreases as the cloud emits, but of course the column of material does not. 
The molecular column densities of positions P1--P10 are $\sim$10$^{22}$\,cm$^{-2}$,
peaking at $\approx$\,7$\times$10$^{22}$\,cm$^{-2}$ for P8 (F07). The latter 
corresponds to 2.8\,mag extinction at 8\,$\mu$m (70\,mag in $A_V$)\footnote{This is a 
factor of 2 above the value in F07, because of a too low conversion factor used by F07.}.
These estimates demonstrate that parts of this cloud have a mid-IR extinction 
comparable to values found for inner Galaxy IRDCs
\citep[e.g.,][]{Carey1998,Hennebelle2001}.

\begin{figure*}[!ht]
\epsscale{1.0}
\plotone{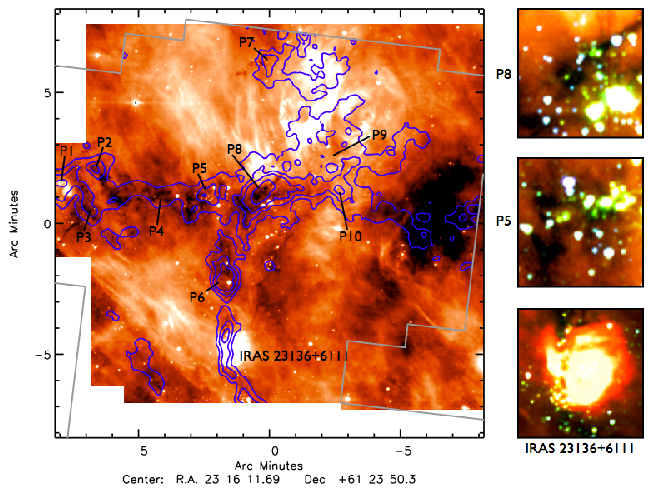}
\caption{This figure shows the IRAC 8\,$\mu$m emission toward
the dark cloud complex G111. The contours depict the integrated 
C$^{18}$O 2-1 emission between -56 and -46\,km\,s$^{-1}$, 
corresponding from outside-in to 2, 3, 4 and 6 K\,km\,s$^{-1}$ (adopted from F07).
The grey line outlines the area covered
by the C$^{18}$O observations. The P-numbers represent target
positions from F07. The right images show a false-color
close-up of positions P8, P5 and IRAS23136+6111. The displayed emission 
is from the four IRAC channels, i.e., 3.6\,$\mu$m (blue), 
4.5\,$\mu$m (green), 5.8\,$\mu$m (yellow) and 8\,$\mu$m (red).
\label{fig1}}
\end{figure*}

\begin{figure*}[!ht]
\epsscale{1.0}
\plotone{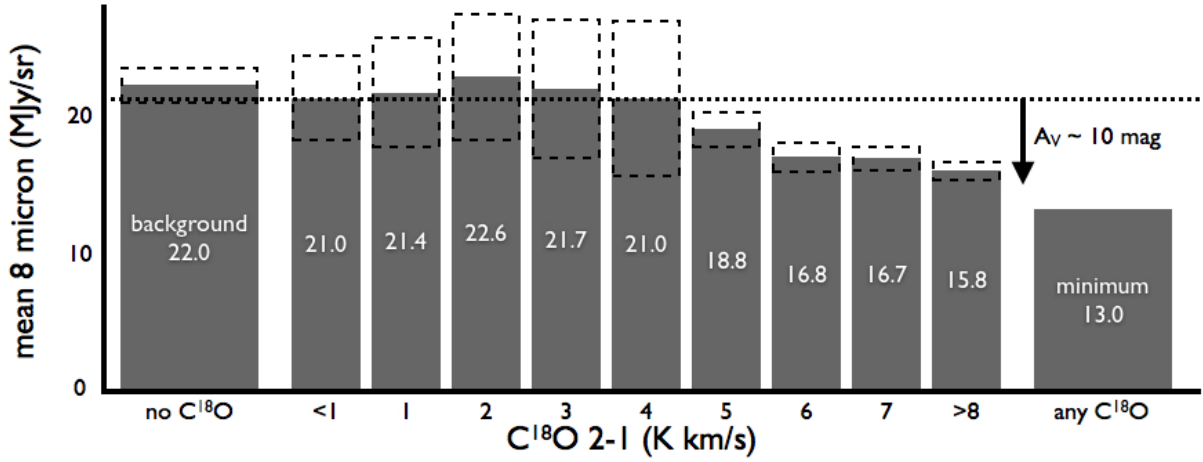}
\caption{Schematic view of the correlation between C$^{18}$O 2-1 and the 8\,$\mu$m diffuse 
emission. The dotted line 
represents a background of 21\,MJy\,sr$^{-1}$. Even though the 
scatter, indicated by the dashed boxes, is significant for low C$^{18}$O emission, 
the trend is clear. The brighter C$^{18}$O emission, i.e., higher column 
densities, corresponds to the lowest mid-IR emission.
\label{fig2}}
\end{figure*}

Recall in this that the selection of IRDCs from {\it MSX} data is based on high contrast 
\citep{Simon2006msx}. In the outer Galaxy, a high contrast is not guaranteed because of the 
low-level background emission. For example, compare the background of the IRDCs toward 
the edge of W51, i.e., $\sim$\,60\,MJy\,sr$^{-1}$ (IRAC/MIPS cycle 1 observations, K. 
Kraemer et al., priv. comm.) with the background that we observe 
($\sim$\,20\,MJy\,sr$^{-1}$). Note that one bright 8\,$\mu$m emission feature clearly is 
in the background of G111, since a dark lane is visible in front of 
IRAS23136+6111 (see Figure \ref{fig1}, lower blow-up).

The four-color images show a clustering of star-like objects near the dark filaments. 
Though these stars could coincidently be in the foreground, this positional consistency 
suggests that some cores might have started to form stars. The excess in steller surface 
density was also pointed out by F07 toward P5 and P8, based on the 2MASS data. 
Note that IRDCs are often associated with active star formation 
\citep[e.g.,][]{Redman2003,Rathborne2005}.
Star forming activity may further reveal itself by the presence of `green fuzzy
emission', i.e., as weak extended 4.5\,$\mu$m features. This emission is often 
attributed to shocked gas arising from outflow-activity, e.g., the pure rotational 
H$_2$ lines S(11) at 4.18\,$\mu$m through S(9) at 4.69\,$\mu$m and the ro-vibrational 
lines of CO at 4.45--4.95\,$\mu$m  \citep[e.g.,][]{Noriega2004,Marston2004}.
Shocked gas features are, besides for nearby star formation, also frequently 
observed toward IRDCs \citep[e.g.,][]{Rathborne2005,Beuther2007}. A first impression of 
the emission at 4.5\,$\mu$m suggests that such features are present toward some cores
in G111. However, further investigation is required to confirm this.
\\ \ \\
\textbf{Notes on individual cores}\\
The positions P1 to P10 in Figure \ref{fig1} were selected by F07 based on their high 
column density. A detailed investigation showed that they represent cold 
(10--20\,K), dense ($>$10$^3$\,cm$^{-3}$) and massive cores 
($\sim$100\,\msun, P8$\sim$\,1000\,\msun) where stars might form. 
Star forming activity was 
investigated by means of 2MASS star-colors and -counts and the presence of warm gas 
(NH$_{3}$, $^{13}$CO). The following discussion summarizes some of the results in
F07 and includes the mid-IR characteristics of the cores presented in this paper.

P1, P2, P3 and P7 do not show indications of active star formation. P1 shows enhanced 
8\,$\mu$m emission, which may be a chance encounter along the line of sight.
The other cores show a decrease in 8\,$\mu$m emission, corresponding to 
an $A_V$ of $>$10\,mag. The peak extinction appears somewhat
offset from the C$^{18}$O peak, which is true for most cores. This may be 
explained by C$^{18}$O freeze-out in the densest regions.
P4 is not a core but part of the C$^{18}$O filament extending to the east. Several 
extinction peaks at 8\,$\mu$m ($A_V$$\gtrsim$10\,mag) are present. 
P9 and P10 are not evident as cores in C$^{18}$O and show no specific features at 
8\,$\mu$m. P6 shows a decrease of 8\,$\mu$m emission ($A_V$$\gtrsim$7\,mag), but the 
contrast with the background may be reduced due to the presence of the bright source 
near the C$^{18}$O peak. Whether this source is physically associated cannot be 
determined at present. P5 and P8 show signs of active star formation. The NH$_{3}$ lines
indicate the presence of warm gas. Both cores have associated 2MASS sources with 
typical near-IR colors of YSOs. Furthermore, the stellar surface 
density of 2MASS peaks at a value almost twice that of the surrounding field for P8. 
These results are supported by the four-color images that reveal signs of active star 
formation in the form of clustering of objects and 4.5\,$\mu$m `green fuzzes'. 
The 8\,$\mu$m contrast indicates peak extinctions of at least 12.5\,mag in $A_V$
for both cores.
IRAS23136+6111 was not a target position in F07, but is nonetheless interesting to include
here. C$^{18}$O emission reveals a narrow filamentary structure. The four-color image 
depicts a dark lane and a clustering of point sources matching the location of the 
filament. We thus conclude that it must be in the foreground of the bright IRAS object.
Further investigation is required to see if the dark lane and the IRAS source are 
associated, which may indicate a triggered star forming event.

To summarize, the {\it Spitzer} observations support the results in F07. 
Except P5 and P8, where star formation may have started, the cores appear to be in 
a cold, pre-stellar phase.
\\ \ \\
\textbf{Future prospects}\\
Supplemental IRAC and MIPS 24\,$\mu$m data from {\it Spitzer} are expected.
Combining the IRAC bands with the 2MASS data enables a 
discrimination between stars and YSOs for many of the clustered point 
sources in the cloud vicinity. Adding 24\,$\mu$m data will allow 
a proper SED modeling of the dust emission. The characteristic features of
envelopes, disks and photospheres in the SED will enable a determination of the 
evolutionary state, i.e., Class 0, I, II or III, of the apparent discrete sources 
\citep[e.g.,][]{Robitaille2007}. Star forming activity can be further 
characterized through an analysis of the spatial distribution and strength 
of polycyclic aromatic hydrocarbon emission (e.g., 3.3, 6.2 and 
7.7\,$\mu$m covered by the 3.6, 5.8 and 8\,$\mu$m bands, respectively) and shock activity, 
exemplified by the 4.5\,$\mu$m features in the blow-up images in Figure 
\ref{fig1} \citep[e.g., bow shocks;][]{Neufeld2006,Velusamy2007}.
Considering the distance to the cloud, high-resolution continuum and spectroscopic 
observations (near-IR to millimeter) are essential to improve on the spatial information.
\acknowledgments
We thank the anonymous referee for his/her careful 
reading of the manuscript and his/her constructive remarks.
{\it Facilities:} \facility{Spitzer (IRAC)}


\clearpage

\end{document}